\documentclass[aps,onecolumn,superscriptaddress,preprintnumbers,amsmath,amssymb,preprint]{revtex4}
\usepackage{graphicx}
\usepackage{amsmath}
\usepackage{hyperref}

\begin{document}

\title{
	Probing Yoctosecond Quantum Dynamics in Toponium Formation at Colliders
}

\author{Chang Xiong}
\affiliation{School of Physics, Beihang University, Beijing 100191, China}

\author{Yu-Jie Zhang}
\thanks{zyj@buaa.edu.cn}
\affiliation{School of Physics, Beihang University, Beijing 100191, China}

\begin{abstract}
	The formation of toponium, a bound state of top and anti-top quarks, provides an unprecedented system for investigating quantum state dynamics at ultrashort timescales. We explore two distinct phenomenological descriptions of this process: a 'wavelike' scenario emphasizing the role of quantum superposition at creation, and a 'particlelike' scenario where a finite formation time is governed by relativistic causality. Both descriptions are compatible with the principles of relativistic quantum field theory. We propose to distinguish these scenarios by exploiting the top quark's intrinsic lifetime ($\tau_t \sim 5.02 \times 10^{-25}$\,s) as a quantum chronometer. We simulate the cross-section ratio $R_b = \sigma(e^+e^- \to b\bar{b})/\sum_q \sigma(e^+e^- \to q\bar{q})$ ($q = u,d,s,c,b$) near $\sqrt{s} = 343$\,GeV at future lepton colliders (CEPC/FCC-ee). These distinct scenarios yield observable $R_b$ profiles, enabling $>5\sigma$ discrimination with 1500\,fb$^{-1}$ of data. Preliminary LHC data provide independent $2$--$3\sigma$ support. This framework establishes a collider-based method to explore time-dependent quantum phenomena in particle production with yoctosecond ($10^{-24}$\,s) resolution.
\end{abstract}

\maketitle

\section*{Introduction}
Understanding the dynamics of quantum state formation is a central theme in modern physics. While spatial aspects of quantum entanglement and nonlocality have been extensively probed through Bell inequality tests~\cite{Bell:1964kc,Aspect:1982fx,ATLAS:2023fsd}, direct experimental investigation of the \emph{temporal structure} of quantum state emergence remains an active frontier. In particular, how quantum superposition manifests and evolves over ultrashort timescales in high-energy processes presents unique opportunities for experimental exploration.

The formation of bound states from highly energetic particles offers a promising avenue for such temporal investigation. Here, we examine the formation of toponium~\cite{Fu:2025yft,CMS:2025kzt}, a bound state of a top quark and an antitop quark, as a prime example. We consider two distinct phenomenological descriptions for this process: a 'wavelike' scenario, where the $t\bar{t}$ pair, created via a pointlike interaction, forms a coherent superposition of bound eigenstates; and a 'particlelike' scenario, where a finite formation time of $t_n \sim n^3 \times 10^{-25}$\,s for the $nS$ states is governed by relativistic causality. Both of these descriptions are compatible with the overarching principles of relativistic quantum field theory, but predict different time-dependent behaviors for the evolving system. Experimentally distinguishing these scenarios can shed light on the dynamics of quantum state formation in real-time.

Toponium provides a unique system for this investigation due to its extreme properties. A $t\bar{t}$ pair is produced via a pointlike interaction at a spatial scale set by the top quark’s Compton wavelength ($\sim 1/(173\,\mathrm{GeV}) \approx 1.14\,\mathrm{am}$). The average separation for the $nS$ state is approximately $11.1\,n^2$\,am, with a characteristic formation time $t_n \sim 9.81 n^3 \times 10^{-26}$\,s. Crucially, the top quark’s intrinsic lifetime, $\tau_t = 5.02 \times 10^{-25}$\,s~\cite{Chen:2023osm}, is comparable to these formation timescales. This makes the top quark's decay a built-in quantum clock, allowing for the real-time resolution of these processes. The distinct time-dependent behaviors predicted by the 'wavelike' and 'particlelike' descriptions can thus lead to measurable experimental signatures.

In particular, the cross-section ratio
\begin{equation}
	R_b = \frac{\sigma(e^+e^- \to b\bar{b})}{\sum_{q=u,d,s,c,b} \sigma(e^+e^- \to q\bar{q})},
\end{equation}
evaluated near the toponium threshold at \(\sqrt{s} \approx 343\,\mathrm{GeV}\), is sensitive to the underlying formation dynamics through interference between resonant and continuum amplitudes~\cite{Fu:2025yft}. The wave and particle scenarios predict distinct energy profiles for \(R_b(\sqrt{s})\), enabling experimental discrimination.

Simulations indicate that future lepton colliders such as CEPC~\cite{CEPCStudyGroup:2023quu} and FCC-ee~\cite{FCC:2018evy}, with 1500\,fb$^{-1}$ of integrated luminosity, can distinguish the two scenarios at $>5\sigma$ significance. Preliminary Large Hadron Collider (LHC) measurements of toponium decay modes further provide $2\sigma$-level support~\cite{Fu:2025zxb}.

This work presents a collider-accessible framework to experimentally probe the time-dependent dynamics of quantum state formation. By leveraging the top quark’s ultrashort decay as a physical quantum clock, we open a new window into exploring the temporal structure of quantum phenomena in particle production.

\section*{Toponium formation}
The interaction between a top quark and an antitop quark is governed by a Coulomb-like QCD potential~\cite{Fu:2025yft,Jiang:2024fyw,Fu:2025zxb}:
\begin{equation}
	V(r) = -\frac{ \lambda}{r}, \quad \lambda = 0.309 \pm 0.010,
\end{equation}
where \(r\) is the interquark separation. The dynamics of the $t\bar{t}$ system are described by the nonrelativistic Hamiltonian~\cite{Sumino:1992ai}:
\begin{equation}
	\hat{H} = 2m_t - i\Gamma_t - \frac{\nabla^2}{m_t} + V(r),
\end{equation}
where \(m_t = 172.57 \pm 0.025\)\,GeV~\cite{Li:2022iav,Fu:2025zxb} is the top quark mass, and \(\Gamma_t = 1.31\)\,GeV~\cite{Chen:2023osm,Chen:2023dsi} is its decay width.

The spin-triplet and spin-singlet $nS$ bound states of the top–antitop system are denoted by \(J_t(nS)\) and \(\eta_t(nS)\), respectively, where \(n = 1, 2, 3, \ldots\) is the principal quantum number~\cite{Fu:2025yft,Jiang:2024fyw}. Key properties of the $nS$ states—including the energy eigenvalue \(E_n\), most probable radius \(r_n\), typical velocity \(v_n\), and wavefunction at the origin \(\psi_{nS}(0)\)—are given by~\cite{dirac1981principles}:
\begin{align}
	E_n &= -\frac{\lambda^2 m_t}{4n^2} = -\frac{4.12}{n^2}\,\mathrm{GeV}, \nonumber\\
	r_n &= \frac{2n^2}{\lambda m_t} = \frac{n^2}{26.7}\,\mathrm{GeV}^{-1}, \nonumber\\
	v_n &= \frac{\lambda}{2n} = \frac{0.155}{n}, \nonumber\\
	|\psi_{nS}(0)|^2 &= \frac{\lambda^3 m_t^3}{8\pi n^3} = \frac{6.03 \times 10^3}{n^3}\,\mathrm{GeV}^3.
\end{align}
The nonrelativistic condition holds well, with \(v_n^2 \approx 0.024/n^2\)~\cite{Bodwin:1994jh}.

The time evolution of the wavefunction includes exponential decay governed by the top quark lifetime $\tau_t$:
\begin{equation}
	\psi(\vec{r}, t) = e^{-i(2m_t + E_n)t} e^{-t/\tau_t} \psi(\vec{r}, 0),
\end{equation}
establishing toponium as a natural yoctosecond-scale chronometer of quantum coherence.

At production ($t = 0^-$), the $t\bar{t}$ pair is created in a pointlike configuration, well approximated by \(\psi(\vec{r}, 0^-) \simeq \delta(\vec{r})\)~\footnote{This approximation is justified by the large top quark width \(\Gamma_t = 1.31\,\mathrm{GeV}\), which keeps its virtuality above \(\sqrt{m_t \Gamma_t}\). This suppresses infrared and Coulomb divergences from soft gluon exchange~\cite{Fadin:1993kt,Denner:1997ia}, rendering higher-order corrections negligible.}. Using the completeness relation~\cite{dirac1981principles},
\begin{equation}
	I = \sum_n |n\rangle\langle n| + \int \mathrm{d}E\, |E\rangle \langle E|,
\end{equation}
the wavefunction immediately after creation becomes:
\begin{equation}
	\psi(\vec{r}, 0^+) = \sum_n \psi_{nS}(0)\, |nS\rangle + \dots,
\end{equation}
where continuum and non-$S$-wave states are omitted. This corresponds to the 'wavelike' scenario, where the initial state at $t=0^+$ is expressed as a coherent superposition of all available bound eigenstates, often associated with an instantaneous formation description in certain models.

Alternatively, in the particle picture~\cite{Bjorken:1989xw,Strassler:1990nw}, the top–antitop pair evolves over finite time into a physical bound state. In the center-of-mass frame, the classical dynamics follow:
\begin{align}
	E_n &= m_t v^2(t) - \frac{\lambda}{2r(t)}, \\
	\frac{dr}{dt} &= \sqrt{ \frac{\lambda}{2m_t r(t)} + \frac{E_n}{m_t} },
\end{align}
with initial condition \(r(0) = 0\) and final radius \(r(t_n) = \langle n| \hat{r}|n \rangle/2 = 3r_n/4\). Solving yields:
\begin{equation}
	t_n = \frac{(4\pi - 3\sqrt{3})\,n^3}{3m_t \lambda^2} \approx 9.81\,n^3 \times 10^{-26}\,\mathrm{s}.
\end{equation}

Relativistic corrections to the kinetic term—replacing \(m_t v^2/2\) with \(m_t(\frac{1}{\sqrt{1 - v^2}} - 1)\)—increase \(t_n\) by only $\sim 2\%$. Boosting to the rest frame of the top quark yields an almost identical value: \(t_n \approx 9.75\,n^3 \times 10^{-26}\,\mathrm{s}\). Modifying the initial separation to \(r(0) = 1/(2m_t)\) instead of zero changes \(t_n\) by merely \(2.35 \times 10^{-27}\,\mathrm{s}\), less than \(1/200\) of \(\tau_t\). These corrections are negligible, validating the classical estimate.

Since \(v(r)\) decreases with increasing separation and becomes imaginary beyond $r = 2r_n$ for bound $nS$ states, the quantum expectation value of formation time exceeds its classical counterpart:
\begin{equation}
	\langle t_n \rangle = \langle n | \int_0^{r} dt | n \rangle > t_n.
\end{equation}

We parametrize the quantum formation time as \(t_n^Q = A t_n\), where
\begin{equation}
	A =
	\begin{cases}
		0 & \text{(wavelike, instantaneous)},\\
		\in [0.5, 2] & \text{(particlelike, causal)}.
	\end{cases}
\end{equation}

This modifies the amplitude by a factor \(e^{-A t_n / \tau_t}\), resulting in cross-section suppression by \(e^{-2A t_n / \tau_t}\).

\begin{figure}[t]
	\centering
	\includegraphics[width=0.85\linewidth]{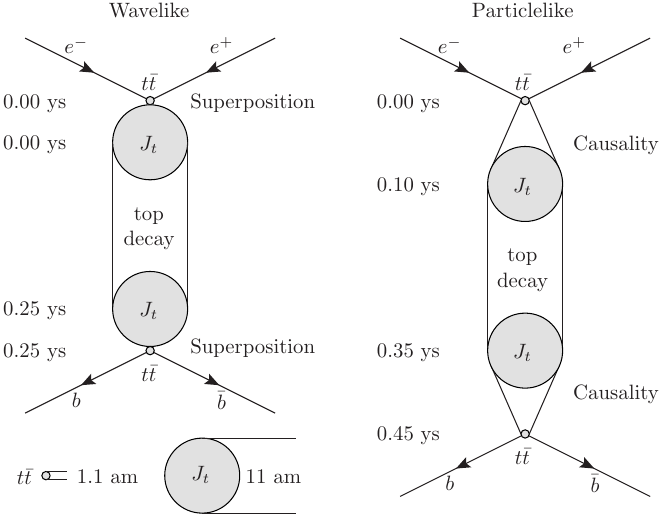}
	\caption{\label{fig:wavelikeparticlelike}
		Time evolution of toponium formation contrasting instantaneous quantum superposition (wavelike) with finite causal propagation (particlelike) in yoctosecond scales.
	}
\end{figure}

The parameter \(A\) captures different phenomenological descriptions of bound-state formation on ultrashort timescales (see Fig.~\ref{fig:wavelikeparticlelike}). The 'wavelike' scenario ($A=0$) describes an instantaneous formation of the superposition of eigenstates. By contrast, the 'particlelike' scenario ($A>0$) adheres to relativistic causality, requiring the top quark pair to take a finite time to form a physical bound state. A value of \(A \approx 1\) corresponds to classical causal evolution under a Coulomb-like potential. In the extreme limit of negligible interactions and light-speed propagation, the time to reach the 1S-state mean separation yields \(A = 0.189\). Although a complete first-principles derivation of \(A\) remains elusive, this parameterization allows for experimental discrimination between these different formation scenarios through measurable suppression in formation amplitudes.

\section*{Collider probes}

The formation dynamics of toponium leave observable imprints in the process \(e^+e^- \to b\bar{b}\) via interference between resonant and continuum amplitudes. The full amplitude is given by~\cite{Fu:2023uzr,Fu:2025zxb,Fu:2025yft}:
\begin{equation}
	\mathcal{M}(e^+e^- \to b\bar{b}) = \mathcal{M}_\gamma + \mathcal{M}_Z + \sum_n \mathcal{M}(J_t(nS)),
\end{equation}
where \(\mathcal{M}(J_t(nS)) \propto e^{-2A t_n / \tau_t}\) includes a dynamical suppression factor that depends on the quantum formation scenario.

To isolate the effect, we define the cross-section ratio~\cite{Fu:2025yft}:
\begin{equation}
	R_b = \frac{\sigma(e^+e^- \to b\bar{b})}{\sum_{q=u,d,s,c,b} \sigma(e^+e^- \to q\bar{q})},
\end{equation}
which is sensitive to interference effects and top-antitop bound state dynamics.

We perform pseudo-experiments to simulate expected measurements at future lepton colliders such as CEPC~\cite{CEPCStudyGroup:2023quu} and FCC-ee~\cite{FCC:2018evy}. The total uncertainty on \(R_b\) at each energy point \(\sqrt{s_i}\) is conservatively estimated as~\cite{Fu:2025yft}:
\begin{equation}
	\Delta R_b^{\mathrm{PEX}}(i) = \pm 1.1 \times 10^{-4}\,\text{(sys)} \pm \frac{83.2 \times 10^{-4}}{\sqrt{\mathcal{L}_i~[\mathrm{fb}^{-1}]}}\,\text{(stat)},
\end{equation}
where \(\mathcal{L}_i\) is the integrated luminosity at point \(i\).
The dominant systematic uncertainties are associated with the potential scheme ($9.8 \times 10^{-5}$) and beam energy spread ($3.6 \times 10^{-5}$). Other sources, including detector efficiency, background subtraction, and luminosity measurement, contribute uncertainties that are an order of magnitude smaller, as detailed in Ref.~\cite{Fu:2025yft}.

We construct a global \(\chi^2\) function incorporating mass constraints:
\begin{equation}
	\label{Eq:chi2IncludeMT}
	\chi^2 = \frac{(m_t - m_t^c)^2}{(\Delta m_t)^2} + \sum_{i=1}^2 \left( \frac{R_b^{\mathrm{PEX}}(i) - R_b^A(i; m_t, \lambda)}{\Delta R_b^{\mathrm{PEX}}(i)} \right)^2,
\end{equation}
where \(m_t^c = 172.57\,\mathrm{GeV}\) is the nominal top mass and \(\Delta m_t = 25\)\,MeV (CEPC)~\cite{Li:2022iav} or \(40\)–\(75\)\,MeV (FCC-ee)~\cite{Schwienhorst:2022yqu}. The number of degrees of freedom is \(\mathrm{ndf} = 2\), corresponding to four data points and two free parameters (\(m_t\) and \(\lambda\)).

\begin{figure}[t]
	\centering
	\includegraphics[width=0.95\linewidth]{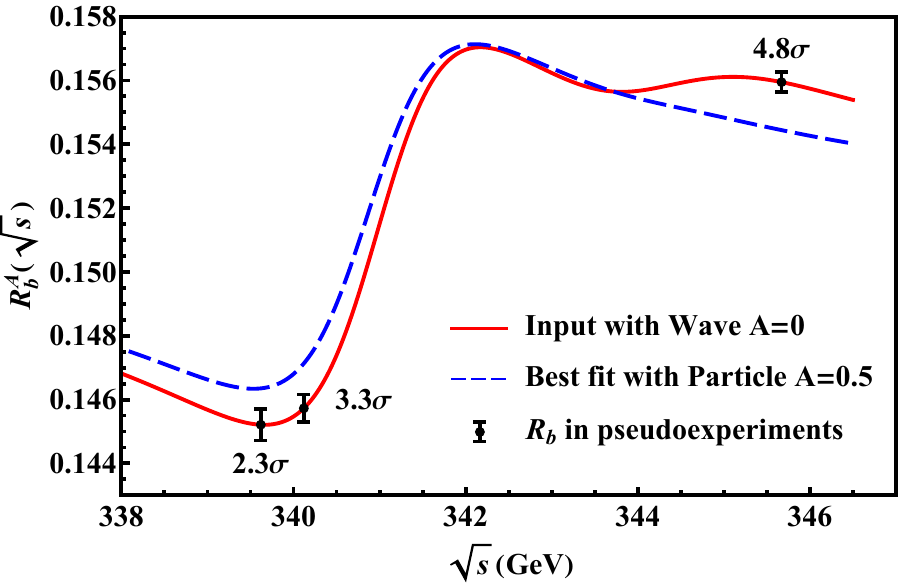}
	\caption{\label{fig:RbInputWave}
		Cross-section ratio \(R_b(\sqrt{s})\) assuming the wave scenario (\(A=0\)). Simulated pseudo-data at three energy points \((339.62,\,340.12,\,345.67)\)\,GeV with integrated luminosities \((300,\,400,\,800)\)\,fb\(^{-1}\) are shown. Red curve: theoretical prediction for wavelike formation (\(A=0\)); blue dashed: best-fit under alternative hypothesis (\(A=0.5\)).
	}
\end{figure}

Assuming the wave scenario (\(A = 0\)) is realized in nature, the alternative particle scenario (\(A = 0.5\)) is excluded at \(6.1\sigma\) significance using luminosities of \((300,\,400,\,800)\)\,fb$^{-1}$ at \(\sqrt{s} = (339.62,\,340.12,\,345.67)\)\,GeV (Fig.~\ref{fig:RbInputWave}). This result is robust under 1$\sigma$ variations in theory parameters: $\lambda = 0.309 \pm 0.010$ ($5.0$–$6.7\sigma$ range) and $m_t = 172.57 \pm 0.025$\,GeV ($5.9$–$6.4\sigma$ range). Even for $\Delta m_t = 40$–$75$\,MeV, the significance remains above $5.1$–$5.8\sigma$. If the mass prior in Eq.~\eqref{Eq:chi2IncludeMT} is dropped, a luminosity of $\sim3000$\,fb$^{-1}$ is needed to reach comparable sensitivity.

Conversely, if the particle scenario (\(A = 0.5\)) is true, the wave scenario (\(A = 0\)) is excluded at \(5.5\sigma\) with \((300,\,475,\,825)\)\,fb$^{-1}$ at \((340.24,\,340.58,\,345.67)\)\,GeV. At $\sqrt{s} = 345.67$\,GeV with 800\,fb$^{-1}$, the predicted values are:
\begin{align}
	R_b^{A=0} &= (1559.5 \pm 2.8_\lambda \pm 0.27_{m_t}) \times 10^{-4}, \nonumber\\
	R_b^{A=0.5} &= (1542.9 \pm 1.5_\lambda \pm 0.27_{m_t}) \times 10^{-4}, \nonumber\\
	\Delta R_b^{\mathrm{PEX}} &= 3.1 \times 10^{-4},
\end{align}
yielding a local significance of \(3.7\sigma\).

In practice, data collection could strategically commence at the third energy point. With 800\,fb$^{-1}$, one can already infer the favored scenario. The remaining luminosity can then be allocated strategically to complete a 1500\,fb$^{-1}$ program, enabling $>5\sigma$ discrimination between instantaneous and causal formation scenarios. Notably, an increase in $A$ leads to a larger suppression of resonant amplitudes, thereby enhancing the difference between the wave and particle scenarios.

\section*{Preliminary test at the LHC}

The LHC offers a complementary platform to probe the temporal dynamics of quantum state formation through decay channels of the spin-singlet toponium states \(\eta_t(nS)\)~\cite{Fu:2025zxb,Han:2024ugl}. In particular, we consider the branching ratio
\begin{equation}
	R_{ZH}^A = \frac{\sigma(pp \to \sum_n \eta_t(nS) \to ZH)}{\sigma(pp \to \sum_n \eta_t(nS) \to t\bar{t})},
\end{equation}
which is sensitive to the parameter \(A\) encoding the temporal formation mechanism.

Assuming degenerate masses and widths across the $nS$ spectrum, the signal rate for \(\eta_t(nS) \to ZH\) scales as \(e^{-4A n^3 t_1 / \tau_t} / n^6\), yielding distinct predictions under the two hypotheses:
\begin{align}
	R_{ZH}^{A=0} &= (18.3 \pm 1.8) \times 10^{-4}, \\
	R_{ZH}^{A=1} &= (8.3 \pm 0.8) \times 10^{-4}.
\end{align}

With approximately 3000\,fb$^{-1}$ of data expected to be collected at the LHC, this branching ratio difference could correspond to a potential \(2\)–\(3\sigma\) level sensitivity~\cite{Yuan:2025,CMS:2025kzt}, offering preliminary evidence in favor of a causal formation delay.

While not definitive, this result provides independent support for the collider-based framework proposed here. It also highlights opportunities for further exploration at the High-Luminosity LHC, where increased statistics and improved systematics may enhance sensitivity to the quantum temporal structure of bound state formation.

\section*{Discussion and outlook}

We have introduced a collider-accessible framework to experimentally investigate the temporal evolution of quantum state formation at the yoctosecond scale (\(10^{-24}\,\mathrm{s}\)). By exploiting the top quark’s ultrashort lifetime as an intrinsic quantum clock, we investigate whether quantum states emerge instantaneously  or evolve over a finite causal time.

This difference in temporal evolution manifests in observable suppression patterns in toponium formation amplitudes. In particular, the cross-section ratio \(R_b\) near threshold at future lepton colliders is shown to provide a clean and statistically significant discriminator. Our simulations indicate that CEPC or FCC-ee, with 1500\,fb$^{-1}$ of data, can distinguish between instantaneous and causal formation scenarios at greater than \(5\sigma\) significance.
Future LHC measurements of $\eta_t(nS)\to ZH$ and $\bar{t}t$ decay modes could provide supporting evidence at the $2$--$3\sigma$ level.

These results provide a unique probe into the applicability of quantum theory in ultrafast, high-energy regimes. Unlike traditional Bell tests, which probe spatial nonlocality, our approach is inherently temporal, providing a direct probe of the temporal unfolding of quantum coherence in bound state formation. This opens a new experimental avenue in the foundations of quantum theory, with far-reaching implications for relativistic quantum information, quantum field theory, and the operational limits of causality.

Future improvements---including the use of polarized beams, improved separation of $b$- and $\bar{b}$-jets, and a better understanding of the potential---will further enhance the sensitivity of this framework. As high-energy colliders advance toward precision quantum tests, they may offer answers to one of the most fundamental questions at the intersection of quantum mechanics and relativity.

\section*{Acknowledgments}
We thank Kuang-Ta Chao, Jing-Hang Fu, Li-Sheng Geng, Yu-Ji Li, Yan-Qing Ma, Ce Meng, Cong-Feng Qiao, and Cheng-Ping Shen for valuable discussions on theoretical modeling and experimental strategies. We also acknowledge Gang Li and Man-Qi Ruan (CEPC), as well as Li Yuan (CMS), for helpful input on collider design and measurements.


\providecommand{\href}[2]{#2}\begingroup\raggedright\endgroup

\end{document}